\documentclass[11pt,a4paper]{article}
\usepackage{amsmath,amssymb,mathrsfs,framed}
\usepackage[colorlinks]{hyperref}
\usepackage{color}
\usepackage[all]{xy}

\newcommand{\dede}[2]{\frac{\delta #1}{\delta #2}}
\newcommand{\curl}{{\operatorname{curl}}}

\newcommand{\rem}[1]{}

\newcommand{\de}{{\rm d}}

\newcommand{\z}{{\mathbf{\hat{z}}}}
\newcommand{\bx}{{\mathbf{x}}}
\newcommand{\bq}{{\mathbf{x}}}

\newcommand{\bp}{{\mathbf{p}}}

\newcommand{\bm}{{\mathbf{m}}}

\newcommand{\bA}{{\mathbf{A}}}

\newcommand{\bJ}{{\mathbf{J}}}
\newcommand{\bK}{{\mathbf{K}}}

\newcommand{\bB}{{\mathbf{B}}}

\newcommand{\bu}{{\boldsymbol{u}}}

\newcommand{\dvol}{{\de^3\bq\,\de^3\bp}}

\begin{document}

\title{Energy-Casimir stability of hybrid Vlasov-MHD models}
\author{Cesare Tronci$^1$, Emanuele Tassi$^{2}$, Philip J. Morrison$^3$
\\
\footnotesize\it $^1$Department of Mathematics, University of Surrey, Guildford GU2 7XH, UK
\\
\footnotesize\it $^2$CNRS \& Centre de Physique Th\'eorique, Campus de Luminy, 
13288 Marseille cedex 9, France\\
\footnotesize\it and Universit\'e de Toulon, CNRS, CPT, UMR 7332, 83957, La Garde, France
\\
\footnotesize\it $^3$Department of Physics \& Institute for Fusion Studies, 
University of Texas, Austin 78712-0262, USA}

\date{}
\maketitle

\bigskip

\begin{abstract} 
Different variants of hybrid kinetic-fluid models are considered for describing the interaction of a bulk fluid plasma obeying MHD and an energetic component obeying a kinetic theory. Upon using the Vlasov kinetic theory for energetic particles, two planar Vlasov-MHD models are compared in terms of their stability properties. This is made possible by the Hamiltonian structures underlying the considered hybrid systems, whose infinite number of invariants makes the energy-Casimir method  effective for determining stability. Equilibrium equations for the models are obtained from a variational principle and in particular a generalized hybrid Grad-Shafranov equation follows for one of the considered models. The stability conditions are then derived and discussed with particular emphasis on kinetic particle effects on classical MHD stability.
\end{abstract}



\bigskip

\section{Introduction}

Several plasma systems  comprise an energetic species interacting with a fluid bulk usually described by magnetohydrodynamics (MHD). For example, in  fusion plasmas, the fusion reactions produce energetic alpha particles that escape the usual fluid closures and require a   kinetic theory description. Another example is provided by the interaction of Earth's magnetosphere with the energetic solar wind.

The interaction of an energetic component with a MHD plasma  can be described by hybrid kinetic-fluid models that couple the energetic  particle kinetics with the MHD equations. The very first nonlinear hybrid models were proposed in the early 90's \cite{Cheng,ParkEtAl}, when it was recognized that two different kinetic-fluid coupling schemes can be consistently formulated. When the coupling occurs by Lorentz force terms in the MHD fluid equation, the hybrid model is known as the  \emph{current-coupling scheme} (CCS); on the other hand, when the coupling occurs by the kinetic  pressure tensor, the corresponding model is referred to as  the \emph{pressure-coupling scheme} (PCS). Moreover,  each scheme may involve different possible variants to describe the energetic  particle kinetics: drift-kinetic, gyro-kinetic and full Vlasov-type equations. While the first two are most used in current computer simulations, the simple mathematical form of the Vlasov equations can be advantageous in analytical studies such as that presented in this work.

Although the PCS is often  preferred over the CCS, the former, in its original formulation, suffers from not conserving the total energy;  consequently,  unphysical energy sources may drive spurious instabilities \cite{TrTaCaMo}. The  formulation of  an energy-conserving PCS variant was done in \cite{Tronci}, upon adopting the  Vlasov description for the energetic  particles. While the fluid momentum equation is unaltered by energy conservation, the accompanying kinetic equation is modified by the presence of terms that arise from expressing the energetic  particle kinetics in the frame of the MHD bulk. The energy-conserving PCS of  \cite{Tronci}  was obtained by applying standard methods in the theory of noncanonical Hamiltonian systems. More particularly, the Hamiltonian structures of MHD \cite{MorrisonGreene,Morrison3,MaWeRaScSp,morrison98,Morrison2005,Morrison2} and of the Vlasov equation \cite{Morrison2bis,Morrison3,MaWe1,MaWeRaScSp,morrison98,Morrison2005,Morrison2} were combined to obtain the Hamiltonian structure of the CCS and of the new Hamiltonian PCS. Then, since the energy-conserving PCS is also Hamiltonian, the general theory offers the opportunity to investigate   stability properties by applying the energy-Casimir method \cite{MoTaTr2014,HoMaRaWe,morrison98},  a method that is a descendent of the Lyapunov method known as  thermodynamic stability in the early plasma literature \cite{KO,fowler,BFHM}. This method has two main advantages:  it is relatively easy to implement and it  gives a stronger notion of stability than that based on linear analysis, for it can lead to mathematically rigorous proofs of  nonlinear stability and nonlinear bounds on growth.   However, the expense paid for this is that  the conditions are generally less definitive.  Also, the energy-Casimir method is  effective when applied to Hamiltonian systems possessing a large number of Casimirs, which is  the case for the CCS and the PCS when the dynamics is restricted to a plane, an approximation that can be relevant for both laboratory and astrophysical plasmas. This paper applies  the  energy-Casimir  method to both the CCS and the Hamiltonian PCS in order to analyze the stability properties of these two systems in planar geometry. 

In the remainder of this Introduction we describe the hybrid models in Sec.~\ref{ssec:eoms} and give a brief description of the energy-Casimir method in Sec.~\ref{ssec:CS}.  Then, in Sec.~\ref{sec:CCS} we consider the CCS.  Here we present the Casimir invariants, obtain  equilibria from the energy-Casimir variational principle, and then proceed to our stability results.  The analogous material is presented in Sec.~\ref{sec:PCS} for the PCS.  Finally, in Sec.~\ref{sec:con} we conclude. 

\subsection{Hybrid Vlasov--MHD in three and two dimensions}
\label{ssec:eoms}

The nonlinear equations of the CCS and PCS appeared first in \cite{ParkEtAl,Cheng}, where they were derived by direct manipulations on the separate equations for MHD and energetic particle kinetics. Later, the CCS and the energy-conserving PCS were derived by operating on the Poisson brackets for MHD and the Vlasov equation \cite{Tronci}. In planar geometry, one can study two possible configurations depending on whether the magnetic field lies on the plane or points vertically. Due to the analogy with reduced MHD \cite{HaHoMaMo,HoMaRaWe}, here we choose to study the first case as it is more physically relevant. This section presents explicitly the nonlinear equations of Vlasov-MHD models in both the CCS and the Hamiltonian PCS variants. 

Upon introducing the Vlasov phase-space density $f(\bq,\bp,t)$  and its first three moments
\[
n=\!\int \!f\,\de^3 \bp
\,,
\qquad\ 
\bK=\!\int\! \bp f\,\de^3 \bp
\,,
\qquad\ 
\Bbb{P}=\!\int\! \bp\bp f\,\de^3 \bp
\,,
\]
(here, $\bp$ is the kinetic momentum coordinate)
the nonlinear equations for the CCS and the Hamiltonian PCS are respectively:
\begin{align}\label{cc-hybrid-momentum}
&\rho\left(\frac{\partial\bu}{\partial t}+\bu\cdot\nabla\bu\right)
= -\nabla\mathsf{p} + \left(q_h n\bu-\,a_h\bK + \mu_0^{-1}\bJ\right)\times\bB\,, 
\\
&
\frac{\partial f}{\partial t}+\frac{\bp}{m_h}\cdot   \nabla f  
+q_h\Big(\frac{\bp}{m_h}-\bu\Big)\times\bB\cdot\frac{\partial f}{\partial \bp}=0\,, 
\label{kinetic-CCS}
\\
&
\frac{\partial\rho}{\partial t} +\nabla\cdot\left(\rho\bu\right)=0\,, 
\qquad\quad\!
\frac{\partial\bB}{\partial t}=\nabla\times\left(\bu\times\bB\right)
\label{cc-hybrid-end}
\,,
\end{align}
and
\begin{align}
&\rho\left(\frac{\partial\bu}{\partial t}+\bu\cdot\nabla\bu\right)=-\nabla{\sf p}-{m_h}^{-1}\nabla\cdot\Bbb{P}
+\mu_0^{-1}\bJ\times\bB
\label{PCS-FuPark2}
\\
\label{hybridMHD2}
&\frac{\partial f}{\partial
t}+\Big(\frac{\bp}{m_h}+\bu\Big)\cdot\nabla f
+\Big[{\bp}\times\big(a_h \bB-\nabla\times\bu\big) 
-\bp\cdot\nabla\bu\Big]\cdot\frac{\partial f}{\partial\bp}=0\,,
\\
\label{hybridMHD3}
&\frac{\partial \rho}{\partial t}+ \nabla\cdot (\rho\bu)=0
\,,\qquad\quad\!
\frac{\partial \bB}{\partial t}=\nabla\times\left(\bu\times\bB\right)
\,,
\end{align}
where $a_h=q_h/m_h$ is the kinetic particle mass-to-charge ratio and we have used the ordinary relation $\bJ=\mu_0^{-1} \nabla\times\bB$. Both systems preserve the total energy
\begin{multline}
H=\frac1{2}\int\!\rho|\bu|^2\de^3\bx+\frac1{2m_h}\int \!f \left|\bp\right|^2\dvol 
 +
 \!\int\! \rho\,\mathcal{U}(\rho)\,\de^3\bx
+\frac1{2\mu_0}\int \left|\bB\right|^2\de^3\bx
\,,
\label{Ham-preMHD}
\end{multline}
where $\mathcal{U}(\rho)$ denotes the MHD internal energy so that the scalar pressure is expressed as $\mathsf{p}=\rho^2  \mathcal{U}' ( \rho)$. The non conservative version of the PCS occurring in the plasma literature is given by \eqref{PCS-FuPark2} and \eqref{hybridMHD3}, while equation \eqref{hybridMHD2} is replaced by \eqref{kinetic-CCS}.

In order to study the simplest case,  while retaining  essential physical features of these models, we  consider the  incompressible limit in two-dimensional geometry. In particular,  we assume that the magnetic and velocity  fields lie in the $x-y$ plane of a Cartesian coordinate system, and we assume translational invariance along the $z$-direction. Consequently, we replace  the first of \eqref{cc-hybrid-end} and \eqref{hybridMHD3} by the divergenceless condition $\nabla\cdot\bu=0$ and  write
\begin{equation}
\bB=\nabla_\perp A\times\z
\,,\qquad \bu=\nabla_\perp \phi\times\z\ \,,\qquad \mathrm{and}\qquad
\omega=\z\cdot\nabla\times \bu=-\Delta \phi\,,
\label{aphi}
\end{equation}
where $A\z$ is the vector potential, $\phi$ is the stream function,  and  $\omega$ is the usual vorticity of the MHD fluid,  the subscript $\perp$ denotes planar components, and $\Delta$ denotes the perpendicular Laplacian without the subscript $\perp$.  Then, after introducing the notation $\widehat{\mathcal{A}}(\bq_\perp)=\int\! {\mathcal{A}}(\bq)\,\de z$, we have the following equations for the planar incompressible CCS:
\begin{align}\label{CCS1-2D}
&\frac{\partial \omega}{\partial t}=[\phi,\omega]+\left[\mu_0^{-1}J-a_h\widehat{K}_z\,,A\right]
\,,\qquad\qquad 
\frac{\partial A}{\partial t} =[\phi,A]
\\\label{CCS2-2D}
&\frac{\partial \hat{f}}{\partial t}+ \frac{\bp_\perp}{m_h}\cdot\nabla_\perp \hat{f} +q_h\Big(\frac{\bp_\perp}{m_h}\cdot\nabla_\perp A+[\phi,A]\Big)\frac{\partial \hat{f}}{\partial p_z}-a_h{p_z}\nabla_\perp A\cdot\frac{\partial \hat{f}}{\partial \bp_\perp}=0
\end{align}
where $[a,b]=\partial_xa\,\partial_yb-\partial_ya\,\partial_xb$ is the canonical Poisson bracket,  $\phi=\left(-\Delta^{-1}\right)\omega$  and $J=-\Delta A$ is the current density, which is directed perpendicular to the $x-y$ plane.

By proceeding analogously, we also find the corresponding equations of motion for the planar incompressible (Hamiltonian) PCS: 
\begin{align}\label{PCS2D-1}
&\frac{\partial \omega}{\partial t}=\left[\phi,\omega\right] +\mu_0^{-1\!}\left[J,A\right] 
+\z\cdot\nabla_\perp\times(\nabla_{\perp\!}\cdot\widehat{\Bbb{P}}_\perp)\,,
\qquad\qquad\qquad \ \  
\frac{\partial A}{\partial t}=[\phi,A]\,,
\\
&\frac{\partial \hat{f}}{\partial t}+\big[\hat{f},\phi\big] +\frac{\bp_\perp}{m_h}\cdot\nabla_\perp \hat{f} -\left(\Big(\frac{\bp_\perp}{m_h}\cdot\z\times\nabla_\perp\Big)\nabla_\perp\phi-\frac{p_z}{m_h}\nabla_\perp A\right)\cdot\frac{\partial \hat{f}}{\partial \bp_\perp}
-\frac{\bp_\perp}{m_h}\cdot\nabla_\perp A\,\frac{\partial \hat{f}}{\partial p_z}=0\,, \label{vla}
\end{align}
where we have defined $\widehat{\Bbb{P}}_\perp=\int\!\bp_\perp\bp_\perp\,\hat{f}\,\de^3\bp$.
The explicit derivation of the above planar models is presented in the Appendix, which makes use of the corresponding Hamiltonian structures. Both models conserve the following expression of the  total energy:
\begin{equation}\label{Ham-preMHD-2D}
H=\frac12\int\!\Big(\omega(-\Delta^{-1})\omega-\frac1{\mu_0}A\,\Delta A+\frac1{m_h}\!\int \!\hat{f}\,|\bp|^2\,\de^3\bp\Big)
\,\de^2\bx\,.
\end{equation}

As outlined in the previous Section, the Hamiltonian nature of the CCS and the PCS presented here allows a Lyapunov stability study that is made possible by direct application of the energy-Casimir method \cite{MoTaTr2014,HoMaRaWe}. This procedure is discussed in the next Section. Unless otherwise specified, the remainder of this paper will refer to the Hamiltonian PCS simply as the PCS and we shall drop the hat symbol for simplicity of notation. Moreover, we shall set all physical constants equal to 1, although we shall restore them in the final stability results. 

\subsection{Stability and the energy-Casimir method}    
\label{ssec:CS}

 Consider a dynamical system $\dot z= V$ (not necessarily Hamiltonian) defined on some space  $\mathcal{Z}$, 
where $z\in\mathcal{Z}$ can be a point or a trajectory, `$\ \dot{\ }\ $' denotes time derivative,  and $V$ is an autonomous 
vector field defined on $\mathcal{Z}$. Equilibria of such system are solutions $z_e$ that satisfy $V(z_e)=0$.  According to the standard definition of stability, an equilibrium point $z_e$ of a dynamical system is said to be stable if, for any neighborhood $N$ of $z_e$ there exists a subneighborhood $S\subset N$ of $z_e$ such that if $\mathring{z}$, an initial condition,  is in $S$ then the trajectory $z(t)\in N$ for all time $t>0$.

 The linear problem associated with the above dynamical system reads $\delta\dot{z}= DV(z_e)\cdot \delta z$  and is obtained by expanding $V(z_e + \delta z)$ to first order.  If  $\delta z$ remains in $N$, then the  system is said to be linearly stable, and to  distinguish this kind of stability from that with dynamics under the full vector field $V$ one adds the adjective nonlinear to describe the latter.

 Assuming a solution of the form $\delta z= \hat{z}\exp(\lambda t)$ the linear problem becomes $(DV-\lambda \, {\rm id})\cdot \hat{z}=0$, where id is the identity operator.  The spectrum of $DV$, 
$\sigma(DV)$,  is the set composed of   $\lambda\in \mathbb{C}$ for which the linear operator $DV-\lambda \, {\rm id}$ 
has no inverse, and an equilibrium point is said to be spectrally stable if, for every $\lambda \in \sigma(DV)$, $\lambda$ lies in the left half complex plane, including the ordinate.   This latter case, in which $\lambda$ is purely imaginary, corresponds to pure oscillations, a case that is sometimes called neutral stability. This is the only kind of spectral stability possessed by Hamiltonian systems. 

 We remark that there are logical relations between linear,  nonlinear,  spectral stability but these can be subtle,  particularly in infinite dimensions.  Some descriptions of implications   between the different kinds of stability can be found, for instance, in 
Refs.\ \cite{morrison98,HoMaRaWe}.  The present work concentrates on the particular kind of  stability  in Hamiltonian systems that we refer to as energy-Casimir stability.

For Hamiltonian systems, the vector field of our dynamical system is generated by a Poisson bracket so that the equations of motion have the form $\dot{z}=\{z,H \}$. In finite dimension, one can write $\dot{z}^i=J^{ij}\partial_jH$, where $J$ is the Poisson bivector (cosymplectic form). Recalling the above definition of equilibrium point, it follows then that, for Hamiltonian systems, equilibrium points satisfy $dH\in {\rm Ker}(J)$.  A consequence of the Poisson bracket identities is the Lie-Darboux theorem, which implies  ${\rm Ker}(J)$ is spanned by Casimir invariants, which satisfy $\{C,f\}=0$, for all functions $f$ (although there are serious unresolved issues with this theorem for infinite-dimensional systems (see, e.g., Ref. \cite{yoshida})). Equilibria of Hamiltonian systems are then critical points of an invariant energy function  
\[
\mathcal{F}= H + C
\] 
and this fact is very useful for establishing stability criteria. 

 For finite-dimensional systems, definiteness (positive or negative) of the quadratic form $\delta^2\mathcal{F}(z_e;\delta z)$ assures both linear and nonlinear stability.  For linear systems $\delta^2\mathcal{F}$ is invariant, in fact it is the Hamiltonian for the linear dynamics,  and its definiteness means the equilibrium point is foliated by nested invariants sets that are topologically spheres. The interior of these sets can thus serve as the subneighborhoods $S$ in the above definition of stability.  For nonlinear systems $\mathcal{F}$ is invariant and, under mild smoothness conditions, $\delta^2\mathcal{F}$ determines the topological character of the level sets of $\mathcal{F}$ in a sufficiently small neighborhood of the equilibrium point $z_e$.  This guarantees that for any neighborhood $N$ there is an $S\subset N$,  determined by some level set of $\mathcal{F}$, within which the flow must remain. 

 For infinite-dimensional systems, such as those of interest here, the situation is considerably more complicated.  In the first place, even if  
$\delta^2\mathcal{F}$ has definite sign, this does not imply that an extremal point that satisfies $\delta\mathcal{F}=0$ is in fact an extremum (maximum or minimum).  Moreover, for both linear and nonlinear stability one requires $\delta^2\mathcal{F}$ to lead to a norm for defining open sets.  This leads to a second and for nonlinear stability an often formidable complication:  even if one is able to construct a norm from $\delta^2\mathcal{F}$, a rigorous proof of nonlinear stability requires that the solution to the dynamical system actually exists in this norm. For most of the infinite-dimensional systems of interest in plasma physics, in particular, global existence results are not available. Even in cases when existence results are available, it  can  turn out that  more than one choice of norm may be constructed   and a given  equilibrium  can be stable in one norm and not in another;  in this  case a physical determination must be made about what is important. The reader interested in seeing what makes up a rigorous application of the energy-Casimir method is referred to Refs.~\cite{Batt,Rein}. 

 In this paper we will derive criteria for  which $\delta^2\mathcal{F}$ is positive definite for our hybrid models, following the practice in the physics literature (e.g.\  \cite{HaHoMaMo,HoMaRaWe,MorEli,morrison98}).   We will refer to this kind of stability as energy-Casimir stability.   We remark that this stability implies linear stability \cite{morrison98,HoMaRaWe} but, for infinite-dimensional systems nonlinear stability is subject to the caveats mentioned above.    
 
Finally, we find it opportune to recall that the classical $\delta W$ criterion conceived and adopted for ideal MHD stability analysis \cite{Ber58} is intimately related to a energy-Casimir stability. Indeed, the variation of the potential energy in terms of the displacement vector, which is denoted as $\delta W$, is related (see, e.g., Ref. \cite{morrison98,HoMaRaWe}) to the second variation $\delta^2\mathcal{F}$, so that  stability in terms of $\delta W$ implies energy-Casimir stability with respect to $\delta^2 \mathcal{F}$. The connection between the two criteria can be extended to reduced models obtained from MHD, as in the case e.g.\ of compressible reduced MHD \cite{Mor13}.

\section{Stability of the planar CCS}
\label{sec:CCS}

This section presents the stability analysis of the planar incompressible CCS model \eqref{CCS2-2D}-\eqref{CCS1-2D}. After reviewing the corresponding Casimirs in both three and two dimensions, these quantities will be used to provide energy-Casimir  stability conditions.

\subsection{Helicity and Casimirs}

In three dimensions, the CCS equations \eqref{cc-hybrid-momentum}-\eqref{cc-hybrid-end} were shown in Ref. \cite{HoTr2011} to  possess  the following conservation law for the cross-helicity invariant: 
\[
\frac{\de}{\de t}\int\bu\cdot\bB\,\de^3\bx=0
\,,
\]
which is accompanied by the usual conservation of magnetic helicity, that is $\int\bB\cdot\bA\,\de^3\bx$. Moreover, collisionless kinetic equations are well known \cite{HoMaRaWe} to possess an infinite family of invariants that are expressed as
\begin{equation}\label{VlasovInv}
C_V[f]=\int\!\Lambda(f)\,\de^3\bx\,\de^3\bp
\,,
\end{equation}
where $\Lambda$ is an arbitrary function of one variable. The above functional  is also preserved by the hybrid CCS.

In planar geometry, the above conservation laws produce a more general family of invariants for the incompressible CCS. This family of invariants can be combined into a single functional
\[
C=C_V[f]+C_\textit{\!MHD}[\omega,A]\,,
\,\qquad\text{ with } \qquad
C_\textit{\!MHD}[\omega,A]=\int\!\Big(\omega\Phi(A)+\Psi(A)\Big) {\de^2 \bx}\,,
\]
where $C_\textit{\!MHD}$ has arbitrary functions $\Phi$ and $\Psi$, and integrations are now over two-dimensional domains.    The Casimir  $C_\textit{\!MHD}$ was obtained in  Ref. \cite{MorHaz} for  ideal MHD in planar geometry, where it was shown that the two terms are  descendent from the  conservation of cross-helicity and magnetic helicity of three-dimensional MHD, respectively.

\subsection{Equilibria via the first variation}

As we shall see, the stability analysis for the planar CCS is actually straightforward due to the direct sum structure of the Poisson bracket underlying CCS \cite{Tronci}.  Upon following the prescriptions in Section \ref{ssec:CS}, the equilibria are identified by the variational principle 
\[
\delta \mathcal{F} =  \delta H+\delta  C_\textit{\!MHD} +\delta C_V=0
\,.
\]
Since the total energy \eqref{Ham-preMHD-2D} can also be split as $H=H_V+H_\textit{\!MHD}$, with
\[
H_\textit{\!MHD}[\omega,A]=\frac12\int\!\Big(\omega(-\Delta^{-1})\omega-A\,\Delta A\Big)\,\de^2\bx\,,
\qquad\quad
H_V[f] = \frac1{2}\int \!f\,|\bp|^2\,\de^2\bx\,\de^3 \bp\,,
\]
the equilibria are found from 
\[
\delta\mathcal{F}_\textit{\!MHD}[\omega,A]+\delta\mathcal{F}_V[f]=0
\,,
\]
where $\mathcal{F}_\textit{\!MHD}= H_\textit{\!MHD} + C_\textit{\!MHD}$ and $\mathcal{F}_V=H_V +C_V$.  Because the Hamiltonian and Casimirs are additive,  this implies 
\[
\delta\mathcal{F}_\textit{\!MHD}[\omega,A]=0
\,,\qquad\qquad\
\delta\mathcal{F}_V[f]=0
\,.
\]
Therefore, the equilibrium states  for the hybrid CCS obtained from the variational principle  are simply given by MHD equilibria coexisting with Vlasov equilibria.  Upon computing $\delta\mathcal{F}_V=\int\!\left(\Lambda'+|\bp|^2/2\right)\delta f d^2\bx d^2\bp$ and 
\[
\delta \mathcal{F}_\textit{\!MHD} =\int\!\big(\Phi(A)-(\Delta^{-1})\omega\big)\delta\omega\, {\de^2 \bx} 
+\int\!\big(-\Delta A+\omega\Phi'(A)+\Psi'(A)\big)\delta A \,  {\de^2 \bx}
\,,
\]
the following equilibrium conditions are obtained:
\begin{eqnarray}
&\phi_e+\Phi(A_e)=0\,,
\label{nla}\\
&\Delta A_e-{\omega}_e\Phi'(A_e)-\Psi'(A_e)=0\,,
\label{gs}
\\
&\frac1{2} {\left|\bp\right|^2}+\Lambda'(f_e)=0\,.
\label{vls}
\end{eqnarray}
where the index $e$ is used to denote quantities in the equilibrium configuration.
All the above equilibrium relations are well known and have been extensively discussed in \cite{HaHoMaMo,HoMaRaWe,Mo86}. In particular, the first equation expresses the fact that, at equilibrium, streamlines are magnetic flux functions. The second equation is a generalization of the  Grad-Shafranov equation that includes flow, while the third implies (upon assuming an invertible $\Lambda'$) isotropic equilibria of the type $f_e=f_e(|\bp|^2/2)$.

In \cite{Mo86} (generalized in \cite{tasso}) it was shown that the transformation
\begin{equation}
\chi(A_e):=\int^{A_e}\!\!\sqrt{1-(\Phi'(A'))^2}\, dA'
\end{equation}
transforms \eqref{nla} and \eqref{gs} into $\Delta \chi=  F(\chi)$, 
where $F= \chi'\, \Phi''\circ A(\chi)$.  Using  \eqref{aphi} and \eqref{nla} gives
\begin{equation}
\Phi'= -\frac{u_e}{B_e}
\label{subA}
\end{equation}
where $u_e=|\bu_e|$ and $B_e=|\bB_e|$, we see that  associated with every sub-Alfv\'enic MHD equilibrium with flow there corresponds a  static equilibrium state.

\subsection{Energy-Casimir stability via the second variation}

The calculation of the second variation $\delta^2\mathcal{F}$ proceeds analogously to those in the  previous section. Indeed, the same arguments lead to the stability criterion
\[
\delta^2\mathcal{F}_\textit{\!MHD}[\omega,A]>0
\qquad\mathrm{and}\qquad
\delta^2\mathcal{F}_V[f]>0
\,.
\]
Then,  stability of the hybrid CCS requires that both the stability conditions for ideal MHD (first inequality) and for the Vlasov kinetic equation (second inequality) are satisfied. More explicitly, upon computing
\begin{eqnarray}
\delta^2 \mathcal{F}_\textit{\!MHD} &=&\int \big( \left|\nabla_\perp \delta \phi -\nabla_\perp \delta\Phi\right|^2  
 +\left(1-(\Phi')^2\right)|\nabla_\perp \delta A|^2 \big) \, \de^2 \bx
 \nonumber\\
&&\hspace{2 cm}
+\int\!\left(\omega\Phi''+\Psi''+\Phi' \Delta\Phi'\right)(\delta A)^2  \, \de^2 \bx\,.
\label{d2F}
\end{eqnarray}
From \eqref{d2F} the following sufficient conditions for positive definiteness  are immediate: 
\begin{align} 
&\vert\Phi'\vert^2< 1\,, \label{sta1-A}
\\
&{\omega}\,\Phi''+\Psi''+\Phi'\Delta \Phi'
>0\,,
\label{sta2-A}
\\
&\Lambda''>0\,.
\label{sta3-A}
\end{align}
In order to add some physical insight to expression \eqref{d2F} and conditions \eqref{sta1-A}--\eqref{sta3-A}, we rewrite  \eqref{d2F} as in Ref. \cite{Mo86}, viz.
\begin{eqnarray}
\delta^2 \mathcal{F}_\textit{\!MHD} &=&\int \Bigg\{
\left|\nabla_\perp \delta \phi + \nabla_\perp \left(\frac{u_e}{B_e}\delta A\right)\right|^2  
 + \left(1- \frac{u_e^2}{B_e^2}\right) |\nabla_\perp \delta A|^2 
 \label{d2Fmo86}\\
&& \hspace{-1 cm} + \ 
 \frac{(\delta A)^2}{B_e^2} \left( 
\left(1- \frac{u_e^2}{B_e^2}\right) \bB_e\times \z\cdot \nabla_\perp  J_{||}
 - \left[ \bB_e\times \z\cdot \nabla_\perp  \frac{u_e^2}{B_e^2}\right]\left[  \bB_e\times \z\cdot \frac{\nabla_\perp  B_e^2}{2B_e^2}
 \right]
\right) \Bigg\} \de^2 \bx\,.
\nonumber
\end{eqnarray}
The first term of \eqref{d2Fmo86} vanishes for Alfv\'enic motions, i.e., motions to neighboring equilibrium states \cite{MorEli}, the second term represents  line bending energy,  reduced by the presence of an equilibrium flow.  The first term of the second line can be recognized as the usual kink term, also modified by the presence of flow, while the last term is interchange-like, that involves curvature and  ram pressure $u_e^2$, taking the role of pressure, which  does not occur in low-$\beta$ reduced MHD. 
 As follows from Eq. \eqref{subA}, the inequality \eqref{sta1-A} implies that one of the conditions for  stability is that the equilibrium flow be sub-Alfv\'enic.  Condition  \eqref{sta2-A} is a sufficient condition for stability, but a more in-depth analysis would show that the line bending term can add a positive contribution.  (See e.g.\ Ref. \cite{ampIIa})   Lastly,  condition \eqref{sta3-A} is the monotonicity-isotropy condition on the distribution function (see e.g.\  \cite{MoPf89,Mo87}).

While the stability analysis of the hybrid CCS does not add any new physical feature to the well-studied stability properties of ideal MHD and the usual convexity relation for Vlasov stability, the hybrid PCS leads to quite different results. These differences are  the subject of the remainder of this paper.

\section{Stability of the planar PCS}
\label{sec:PCS}

This section contains the stability analysis of the planar PCS model \eqref{PCS2D-1} and \eqref{vla}. After reviewing the corresponding Casimirs in both three and two dimensions, these quantities will be used to provide stability conditions.  Although the discussion of Casimirs and equilibria for this case can also be found  in Ref. \cite{MoTaTr2014}, here we present new stability conditions that recover the usual condition \eqref{sta3-A} for noncollisional kinetics.

\subsection{Helicity and Casimirs}

In three dimensions, the hybrid PCS  of \eqref{PCS-FuPark2} and \eqref{hybridMHD3} preserves the usual magnetic helicity as well as the Vlasov invariant \eqref{VlasovInv}. On the other hand, the usual expression for  the cross-helicity is not conserved and a velocity shift is required for the conservation of a modified cross-helicity, which reads \cite{HoTr2011}
\[
\frac{\de}{\de t}\!\int\bB\cdot(\bu+\rho^{-1}\bK)\, \de^3\bx=0.
\]

Then, in two dimensions, one obtains the following  infinite set of Casimir invariants for the incompressible PCS:
\[
C=C_V[f]+C_\textit{\!MHD}[\omega,A]-\int\!\Phi(A)\,\z\cdot\nabla\times\bK_\perp  \,{\de^2 \bx}
\]
where $\bK_\perp:=\int\!\bp_\perp f\,\de^3\bp$.
The proof of the conservation relation $\dot{C}=0$ was presented in Ref. \cite{MoTaTr2014}, by using the Poisson bracket structure. The last term above  couples  the parallel vector potential $A$ to the Vlasov distribution $f$ and, as we shall see, this has a substantial effect on the equilibrium configurations and their stability conditions.

\subsection{Equilibria and hybrid Grad-Shafranov relation}

Consider  now the equilibrium conditions for the incompressible PCS in planar geometry.
Following again the general theory of  Sec.~\ref{ssec:CS}, we set  $\delta\mathcal{F}=0$.  
Unlike the CCS, the equilibrium condition for the PCS do not split. They read
\[
\delta\mathcal{F}_\textit{\!MHD}[\omega,A]+\delta\mathcal{F}_V[f] +\delta\left[\int\!  {\de^2 \bx} \,  \Phi(A)\, \z\cdot\!\!\int\!\bp_\perp \times \nabla_\perp f\,\de^3\bp\right] =0
\,. 
\]
Upon computing the variation of the new term, 
\begin{eqnarray}
\delta\left[\int\!   {\de^2 \bx}\,  \Phi(A)\, \z\cdot\!\!\int\!\bp_\perp \times \nabla_\perp f\,\de^3\bp\right] &=&
\int\! {\de^2 \bx}\,  \delta A \Phi'\, \z\cdot\!\!\int\!\bp_\perp \times \nabla_\perp f\,\de^3\bp
\\
&&
\hspace{1 cm} -\!\iint\!\delta f\,\z\cdot\bp_\perp\times\nabla_\perp \Phi\ \de^2\bx\,\de^3\bp\,,
\nonumber
\end{eqnarray}
the equilibrium conditions are seen to be 
\begin{align}
&\phi_e+\Phi(A_e)=0
\label{pcs1}
\\
&-\Delta A_e+\left(\omega_e-\z\cdot\nabla_\perp\times\bK_\perp\right)\Phi'(A_e)+\Psi'(A_e)=0
\label{pcs2}
\\
&\z\cdot\nabla_\perp \Phi(A_e)\times\bp_\perp+\frac{\left|\bp\right|^2}{2}+\Lambda'(f_e)=0
\,.
\label{pcs3}
\end{align}
Using   $\nabla_\perp\phi_e=\z\times\bu_e$, \eqref {pcs1} and \eqref{pcs3} imply
\begin{align}\label{VlasEqRel}
\frac12\left|\bp+\bu_e\right|^2-\frac12\left|\bu_e\right|^2+\Lambda'
= 0
\,,
\end{align}
so that assuming an invertible $\Lambda'$, we obtain the following class of anisotropic equilibria: 
\begin{align*}
f_e=&f_e\!\left(\frac12\left|\bp+\bu_e\right|^2-\frac12\left|\bu_e\right|^2\right)\,.
\end{align*}
Next, an explicit computation \cite{MoTaTr2014} yields
\[
\z\cdot\nabla\times\!\int\!\bp_\perp\,f_e\,\de^3\bp=-\nabla_\perp\cdot\left(n\nabla_\perp\Phi\right)=-\z\cdot\nabla_\perp\times(n\bu_e)
\]
so that Eq. \eqref{pcs2} reads
\begin{equation}
-\Delta A_e+\Phi'(A_e)\,\nabla_\perp\cdot\big((1+n_e)\nabla_\perp\Phi(A_e)\big)+\Psi'(A_e)=0\,,
\label{hGS}
\end{equation}
which we shall call the  \emph{hybrid Grad-Shafranov equation}.   Recognize that the special case $\Phi(A_e)=-A_e$ (Alfv\'en equilibrium) takes the above relation to
\[
\nabla_\perp\cdot\left(n_e\nabla_\perp A_e\right)+\Psi'(A_e)=\z\cdot\nabla_\perp\times(n_e\bu_e)+\Psi'(A_e)=0
\quad\implies\quad
A_e=A_e\big(\z\cdot\nabla_\perp\times(n_e\bu_e)\big)
\,,
\]
while  the case $\Phi(A_e)=0$ produces the usual  static Grad-Shafranov equilibria of ideal MHD \cite{HaHoMaMo,HoMaRaWe}.

Having characterized the  equilibria of the planar PCS, we can consider  the energy-Casimir  conditions  for stability, which we do in the next section.

\subsection{Stability conditions and kinetic particle effects}

As before, we  compute the second variation, giving 
\begin{align*}
\delta^2\mathcal{F}=
&\ \delta^2 \mathcal{F}_\textit{\!MHD} +\iint\Lambda''(\delta f)^2  \de^2 \bx \de^3 \bp +\!\iint\!\left(\z \cdot \bp_\perp \times \nabla_\perp  f \right)\Phi''(A) (\delta A)^2  {\de^2 \bx \de^3 \bp}
\\
&-2\iint\!\Lambda''\left(\frac1{\Lambda''}\,\delta f\,(\nabla_\perp \Phi'\,\delta A+\Phi'\,\nabla_\perp\delta A)\cdot\z\times \bp_\perp\right)  {\de^2 \bx \de^3 \bp}
\\
=& \ 
\delta^2 \mathcal{F}_\textit{\!MHD}+\iint\!\left(\z \cdot \bp_\perp \times \nabla_\perp f  \right)\Phi''(A) (\delta A)^2  {\de^2 \bx \de^3 \bp}
\\
&+\frac12\iint\!\Lambda''\,\Big|\delta f-\frac2{\Lambda''}\,\nabla_\perp \Phi'\delta A\cdot\z\times \bp_\perp\Big|^2  {\de^2 \bx \de^3 \bp}
\\
&
+\frac12\iint\!\Lambda''\,\Big|\delta f-\frac2{\Lambda''}\,\Phi'\nabla_\perp\delta A\cdot\z\times \bp_\perp\Big|^2  {\de^2 \bx \de^3 \bp}
\\
&
-2\iint\!\frac1{\Lambda''}\Big| \bp_\perp\times\nabla_\perp \Phi'\delta A\Big|^2  {\de^2 \bx \de^3 \bp}
-2\iint\!\frac1{\Lambda''}\Big|\Phi' \bp_\perp\times\nabla_\perp\delta A\Big|^2  {\de^2 \bx \de^3 \bp}
\\
=&\ 
\delta^2 \mathcal{F}_\textit{\!MHD} +\iint\!\left(\z \cdot \bp_\perp \times \nabla_{\perp}f \right)\Phi''(A) (\delta A)^2  {\de^2 \bx \de^3 \bp}
\\
&+\frac12\iint\!\Lambda''\,\Big|\delta f-\frac2{\Lambda''}\,\nabla_\perp \Phi'\delta A\cdot\z\times \bp_\perp\Big|^2  {\de^2 \bx \de^3 \bp}
\\
&+\frac12\iint\!\Lambda''\,\Big|\delta f-\frac2{\Lambda''}\,\Phi'\nabla_\perp\delta A\cdot\z\times \bp_\perp\Big|^2  {\de^2 \bx \de^3 \bp}
\\
&
-2 {\int}\!(\operatorname{Tr}\Pi_\perp)|\nabla_\perp\Phi'|^2(\delta A)^2  {\de^2 \bx}+2\iint\!\frac1{\Lambda''}\,(\bp_\perp\cdot\nabla_\perp\Phi')^2(\delta A)^2  {\de^2 \bx \de^3 \bp}
\\
&
-2 {\int} \!\Phi'^2(\operatorname{Tr}\Pi_\perp)|\nabla_\perp\delta A|^2  {\de^2 \bx}
+2
\iint\!\frac1{\Lambda''}\,\Phi'^2(\bp_\perp\cdot\nabla_\perp\delta A)^2  {\de^2 \bx \de^3 \bp}
\end{align*}
where  $\Pi_\perp:=\int\! (\Lambda'')^{-1}\bp_\perp\bp_\perp\, \de^3\bp$. 
Thus, the PCS stability conditions are
\begin{align} 
&\vert\Phi'\vert^2< \frac{1}{1+2\operatorname{Tr}\Pi_\perp} 
\label{sta1}
\\
&\overline{\omega}\,\Phi''+\Psi''+\Phi'\Delta \Phi'
-2|\nabla_\perp\Phi'|^2\operatorname{Tr}\Pi_\perp
>0
\label{sta2}
\\
&\Lambda''>0,
\label{sta3}
\end{align}
where $\overline{\omega}=\omega - \z \cdot \nabla_\perp \times  \bK$ is a shifted vorticity accounting for the contributions of both the MHD and the kinetic components.

Notice that $\phi_e=-\Phi(A_e)$ implies $\bu_e=-\Phi'(A_e)\bB_e$, so that the first stability condition reads
\[
B_e^2\geq\left(1+2\operatorname{Tr}\Pi_\perp\right)u_e^2
\,.
\]

\noindent
It follows from this expression that, for instance in the presence of a Maxwellian equilibrium distribution function, a condition stronger than (\ref{sta1-A}) is required on the equilibrium flow for stability. Compared to the case of reduced MHD or to the CCS, it turns out indeed, that a sub-Alfv\'enic equilibrium flow is no longer sufficient for stability in the presence of a population of kinetic particles, even if the latter follows a Maxwellian distribution.

The condition (\ref{sta2}), on the other hand, can be reformulated as 
\begin{multline}   
\frac{1}{B_e^2} \left[
\left(1- \frac{u_e^2}{B_e^2}\right) \bB_e\times \z\cdot \nabla_\perp  J_{||}
 - \left( \bB_e\times \z\cdot \nabla_\perp  \frac{u_e^2}{B_e^2}\right)\left(  \bB_e\times \z\cdot \frac{\nabla_\perp  B_e^2}{2B_e^2}
 \right)
\right]  \\
+\frac{u_e}{B_e}\frac{\bB_e \times \z}{B_e^2}\cdot \nabla_{\perp} ( \z \cdot \nabla \times \bK)- 2 \left| \nabla_\perp \left( \frac{u_e}{B_e} \right)\right|^2 \operatorname{Tr}\Pi_\perp  >0.\quad
\label{sta2bis}
\end{multline}

In the first line of the inequality (\ref{sta2bis}) one can recognize the terms originated from $\delta^2 \mathcal{F}_{MHD}$ and already discussed in the case of the CCS. The remaining terms, on the other hand, are peculiar to the PCS stability conditions and show an influence of the presence of the kinetic population. In particular, in the first term of the second line, one is able to identify a further possible source of instability analogous to the kink instability but with the role of the current density played by the vorticity associated with the mean flow of the kinetic species. The last term, on the other hand is destabilizing and depends on the second order moments of the equilibrium distribution function.

Finally, upon deriving the equilibrium relation \eqref{VlasEqRel} with respect to $\bp$, we obtain
\[
\Lambda''=-\frac1{f'_e}\ \Longrightarrow\ {f'_e}<0
\qquad\text{and}\qquad
\Pi_\perp=-\!\int\! {f'_e}\ \bp_\perp\bp_\perp\, \de^3\bp
\,,
\]
which states that distributions such that $f_e'=-\nu^2f_e$ (for some constant $\nu$, e.g. the Gaussian distribution) are stable equilibria.

\section{Conclusions}
\label{sec:con}

The Hamiltonian structures of two hybrid kinetic-MHD models have been presented and their energy-Casimir  stability properties have been analyzed. The two adopted models are examples of the two paradigmatic coupling schemes adopted for hybrid models, i.e.,  the CCS and the PCS. For the latter, in particular, the model we investigated corresponds to a modification of the one derived in Ref. \cite{Cheng}, in such a way that the resulting hybrid model is energy-conserving and possesses a Hamiltonian structure. The Casimir invariants for such models have been described and compared with those of ideal MHD in two and three dimensions. 
Equilibria obtained by extremizing $\delta \mathcal{F}$ and the stability conditions corresponding to the definiteness of $\delta^2 \mathcal{F}$ have then been derived and discussed. 

In the case of the CCS, equilibrium relations and stability conditions derived with the energy-Casimir method correspond to those that one would obtain considering independently 2D planar MHD and the Vlasov equation. In particular, we identified (by imposing definiteness of $\delta^2 \mathcal{F}$) the conditions of having a sub-Alfv\'enic MHD flow, the convexity of the distribution function, as well as the stability with respect to the ideal kink instability and to an interchange-like instability. 

A more interesting situation occurs for the PCS. In this case, setting to zero the first variation of the free energy functional yields, among the equilibrium equations, a generalized hybrid Grad-Shafranov equation, which accounts for the presence of both the MHD flow and the kinetic  particle population. With regard to stability, it emerges that the presence of a kinetic particle population leads to a different condition on the MHD equilibrium speed, whereas the convexity condition on the distribution function remains valid. In the case of a Maxwellian distribution for the kinetic species, the MHD flow has indeed an upper bound corresponding to the Alfv\'en speed diminished by a quantity proportional to the equilibrium pressure of the kinetic population. 

As complement to the above results, we have  provided an  Appendix containing  derivations  of the planar CCS and PCS models and of their Hamiltonian structures.

\bigskip

\paragraph{\large Acknowledgements.\!\!} CT acknowledges partial support by the Institute of Mathematics and its
Applications Grant No. SGS27/13 and by the London Mathematical Society Grants No. 31320 \& 41371. ET  acknowledges financial support
received from the Agence Nationale de la Recherche (ANR GYPSI n. 2010 BLAN 941 03) and from the CNRS through the PEPS project GEOPLASMA2. PJM was supported by U.S. Dept. of Energy Contract \# DE-FG05-80ET-53088.
 
\bigskip

\appendix

\section{Explicit derivation of planar Vlasov-MHD}
This Appendix presents the explicit derivation of both CCS and Hamiltonian PCS in planar geometry.

\subsection{Planar current-coupling scheme}

Consider the incompressible CCS equations \eqref{cc-hybrid-momentum}-\eqref{cc-hybrid-end}. These comprise a Hamiltonian system with Hamiltonian \eqref{Ham-preMHD} and Poisson bracket \cite{Tronci}
\begin{align}
\label{PB-current-hybridMHD}
&\{F,G\}_{CCS}= \int{\bm}\cdot \left(\frac{\delta
F}{\delta{\bm}}\cdot\nabla\frac{\delta G}{\delta
{\bm}}-\frac{\delta
G}{\delta{\bm}}\cdot\nabla\frac{\delta F}{\delta
{\bm}}\right)\de^3\bq
\nonumber\\
&\hspace{.5 cm}+
\int \!f\bigg(\left\{\frac{\delta
F}{\delta f},\frac{\delta G}{\delta f}\right\} 
+q_h\,\bB\cdot\frac{\partial}{\partial \bp}\frac{\delta
F}{\delta f}
\times
\frac{\partial}{\partial \bp}\frac{\delta
G}{\delta f}\bigg)\dvol
\nonumber\\
&\hspace{.5 cm}+
q_h
\int\! f \,\bB\cdot\bigg(\frac{\delta F}{\delta \bm}\times
\frac{\delta G}{\delta \bm} 
-\frac{\delta F}{\delta \bm}
\times
\frac{\partial}{\partial \bp}\frac{\delta
G}{\delta f}+\frac{\delta G}{\delta \bm}
\times
\frac{\partial}{\partial \bp}\frac{\delta
F}{\delta f}\bigg)\dvol
\nonumber\\
&\hspace{.5 cm}
+
\int \bB\cdot\left(\frac{\delta F}{\delta \bm}\times\nabla\times\frac{\delta G}{\delta \bB}-\frac{\delta G}{\delta \bm}\times\nabla\times\frac{\delta F}{\delta \bB}\right)\de^3\bq
\nonumber\\
&\hspace{.5 cm}
- \int \!\rho \left(\frac{\delta F}{\delta \bm}\cdot\nabla\frac{\delta G}{\delta \rho} 
-\frac{\delta G}{\delta \bm}\cdot \nabla\frac{\delta F}{\delta \rho}\right)\de^3\bq
\,,
\end{align}
where $\bm= \rho \bu$ is the fluid momentum, as it is expressed in terms of the velocity $\bu$ and the density $\rho$.
Then, we use the formula $\bB=\nabla\times(A\z)=\nabla_\perp A\times\z$ together with the variational relation
\[
\dede{F}{\bm}=\nabla\times\left(\dede{F}{\omega}\z\right)=\nabla_\perp\dede{F}{\omega}\times\z
\]
where $\omega=\z\cdot\nabla\times\,\bm$.  (See e.g.\  \cite{amp0,amp1} for discussion of such relations.)  With these formulas in mind we compute the following term appearing in the fourth line of \eqref{PB-current-hybridMHD}:
\begin{align*}
\int\!\bB\cdot\left(\dede{F}{\bm}\times\nabla\times\dede{G}{\bB}\right) {\de^3 \bx}=&\int\!\curl(A\z)\cdot\left(\curl\left(\dede{F}{\omega}\z\right)\times\dede{G}{\bA}\right){\de^3 \bx}
\\
=-&\int\!\dede{G}{\bA}\cdot\left(\curl\left(\dede{F}{\omega}\z\right)\times\curl(A\z)\right){\de^3 \bx}
\\
=-&\int\!A\left[\dede{F}{\omega},\dede{G}{A}\right]{\de^3 \bx}
\end{align*}
where we have used $\delta G/\delta A=\z\cdot\delta G/\delta \bA$ and we have introduced the planar bracket $[k,h]=\partial_x k\partial_y h -\partial_y k\partial_x h$.
More importantly, we have
\begin{align*}
\bB\cdot\left(\dede{F}{\bm}\times\dede{G}{\bm}\right)&=0
\end{align*}
and
\begin{align*}
\int\!f\,\bB\cdot\dede{F}{\bm}\times\frac{\partial}{\partial \bp}\dede{G}{f} {\de^3 \bx \de^3 \bp}
=&-\int\!f \frac{\partial}{\partial p_z}\dede{G}{f}\,\nabla_\perp\dede{F}{\omega}\cdot\nabla_\perp A\times\z {\de^3 \bx}
\\
=&-\int\!f \frac{\partial}{\partial p_z}\dede{G}{f}\,\z\cdot\nabla_\perp\dede{F}{\omega}\times\nabla_\perp A {\de^3 \bx}
\\
=&
\int\!f \,\frac{\partial}{\partial p_z}\dede{G}{f}\left[A,\dede{F}{\omega}\right] {\de^3 \bx}
.
\end{align*}
In addition, one computes
\begin{align*}
\int\!f\,\bB\cdot\frac{\partial}{\partial\bp}\dede{F}{f}\times\frac{\partial}{\partial\bp}\dede{G}{f} {\de^3 \bx}=
&
\int\!f\,\nabla_\perp A\cdot\z\times\frac{\partial}{\partial\bp}\dede{F}{f}\times\frac{\partial}{\partial\bp}\dede{G}{f}{\de^3 \bx}
\\
=&\int\!f\,\nabla_\perp A\cdot\left(\frac{\partial}{\partial p_z}\dede{G}{f}\frac{\partial}{\partial \bp_\perp}\dede{F}{f}-\frac{\partial}{\partial p_z}\dede{F}{f}\frac{\partial}{\partial \bp_\perp}\dede{G}{f}\right){\de^3 \bx}
\end{align*}
Thus, replacing the above results in \eqref {PB-current-hybridMHD} yields the following  Poisson bracket of the planar incompressible CCS:\begin{align}\nonumber
\{F,G\}=&\int\omega\left[\dede{F}{\omega},\dede{G}{\omega}\right] {\de^3 \bx}+\int A\left(\left[\dede{F}{\omega},\dede{G}{A}\right]-\left[\dede{G}{\omega},\dede{F}{A}\right]\right){\de^3 \bx}
\\\nonumber
&+\int f\left(\frac{\partial}{\partial v_z}\dede{G}{f}\left[\dede{F}{\omega},A\right]-\frac{\partial}{\partial v_z}\dede{F}{f}\left[\dede{G}{\omega},A\right]\right) {\de^3 \bx \de^3 \bp}
\\\nonumber
&+\int\!f\left(\left\{\dede{F}{f},\dede{G}{f}\right\}+\nabla_\perp A\cdot\!\left(\frac{\partial}{\partial p_z}\dede{G}{f}\frac{\partial}{\partial \bp_\perp}\dede{F}{f}-\frac{\partial}{\partial p_z}\dede{F}{f}\frac{\partial}{\partial \bp_\perp}\dede{G}{f}\right)\right) {\de^3 \bx \de^3 \bp}
\end{align}
Then, upon using the total energy \eqref{Ham-preMHD-2D},
one obtains 
\begin{align*}
\partial_t\omega=&[\phi,\omega]+\left[J-\int\!p_z\,f\,\de^3\bp\,\de z\,,A\right]
\,,\qquad
\partial_t A=[\phi,A]
\\
\partial_t f+&\bp\cdot\frac{\partial f}{\partial\bq}+\Big(\bp_\perp\cdot\nabla_\perp A+[\phi,A]\Big)\frac{\partial f}{\partial v_z}-v_z\nabla_\perp A\frac{\partial f}{\partial \bp_\perp}=0
\,.
\end{align*}
Notice that the longitudinal coordinate $z$ plays no role in the dynamics and thus it can be integrated out to yield the equations \eqref{CCS1-2D}-\eqref{CCS2-2D}.

\subsection{Planar pressure-coupling scheme}
Consider the incompressible PCS equations \eqref{PCS-FuPark2}-\eqref{hybridMHD3}. These comprise a Hamiltonian system with Hamiltonian \eqref{Ham-preMHD} and Poisson bracket \cite{Tronci}
\begin{align}
&\{F,G\}_{\tiny PCS}= \int{\bm}\cdot \left(\frac{\delta
F}{\delta{\bm}}\cdot\nabla\frac{\delta G}{\delta
{\bm}}-\frac{\delta
G}{\delta{\bm}}\cdot\nabla\frac{\delta F}{\delta
{\bm}}\right)\de^3\bq
\nonumber
\\
&
\hspace{.5cm}+ \int f\left(\bigg\{\frac{\delta
F}{\delta f},\frac{\delta G}{\delta f}\right\}
 +q_h\, \bB\cdot\frac{\partial}{\partial \bp}\frac{\delta
F}{\delta f}
\times
\frac{\partial}{\partial \bp}\frac{\delta
G}{\delta f}
\bigg)\dvol
\nonumber\\
&\hspace{.5 cm}
+ \int f\left(\left\{\frac{\delta
F}{\delta f},\bp\cdot\frac{\delta G}{\delta \bm}\right\}-\left\{\frac{\delta
G}{\delta f},\bp\cdot\frac{\delta F}{\delta \bm}\right\}\right)\dvol
\nonumber\\
&\hspace{.5 cm}
+ \int \bB\cdot\left(\frac{\delta F}{\delta \bm}\times\nabla\times\frac{\delta G}{\delta \bB}-\frac{\delta G}{\delta \bm}\times\nabla\times\frac{\delta F}{\delta \bB}\right)\de^3\bq
\nonumber\\
&\hspace{.5 cm}
 -\int \rho \left(\frac{\delta F}{\delta \bm}\cdot\nabla\frac{\delta G}{\delta \rho} 
-\frac{\delta G}{\delta \bm}\cdot \nabla\frac{\delta F}{\delta \rho}\right)\de^3\bq
\,.
 \label{PB-pressure-hybridMHD}
\end{align}

In 2D, the terms in the third row become
\begin{align}
\int\!f\left\{\dede{F}{f},\bp\cdot\dede{G}{\bm}\right\}\dvol
&=\int\!f\left\{\dede{F}{f},\bp\cdot\nabla_\perp\dede{G}{\omega}\times\z\right\}\dvol
\\
&=\int\!\dede{G}{\omega}\,\z\cdot\nabla\times\!\int\!\bp_\perp\!\left\{\dede{F}{f},f\right\}\dvol
\\
&=\int\!f\left\{\dede{F}{f},\z\cdot\bp_\perp\times\nabla_\perp\dede{G}{\omega}\right\}\dvol
.
\end{align}
The remaining terms in the bracket are computed as in the previous section of this Appendix. 
Then, the Poisson bracket for the incompressible hybrid PCS reads as
\begin{align}
\{F,G\}=&\int\omega\left[\dede{F}{\omega},\dede{G}{\omega}\right]{\de^3 \bx}+\int A\left(\left[\dede{F}{\omega},\dede{G}{A}\right]-\left[\dede{G}{\omega},\dede{F}{A}\right]\right){\de^3 \bx}
\\
&+\int f\left(\left\{\dede{F}{f},\z\cdot\bp_\perp\times\nabla_\perp\dede{G}{\omega}\right\}-\left\{\dede{G}{f},\z\cdot\bp_\perp\times\nabla_\perp\dede{F}{\omega}\right\}\right){\de^3 \bx \de^3 \bp}
\\
&+\int\!f\left(\left\{\dede{F}{f},\dede{G}{f}\right\}+\nabla_\perp A\cdot\!\left(\frac{\partial}{\partial v_z}\dede{G}{f}\frac{\partial}{\partial \bp_\perp}\dede{F}{f}-\frac{\partial}{\partial v_z}\dede{F}{f}\frac{\partial}{\partial \bp_\perp}\dede{G}{f}\right)\right){\de^3 \bx \de^3 \bp}
\end{align}
Then, upon using the total energy \eqref{Ham-preMHD-2D},
one obtains 
\begin{align}
&\partial_t\omega=\left[\psi,\omega\right]+\left[J,A\right]+\z\cdot\nabla\times\left(\nabla_\perp\cdot\!\int\!\bp_\perp\bp_\perp\,\varphi\,\de^3\bp\right)\,,\qquad
\partial_t A=[\psi,A]
\\
&\frac{\partial f}{\partial t}+\left[f,\psi\right]+\bp\cdot\frac{\partial f}{\partial\bq_\perp}-\left((\z\cdot\bp_\perp\times\nabla_\perp)\nabla_\perp\psi-v_z\nabla_\perp A\right)\cdot\frac{\partial f }{\partial \bp_\perp}-\bp_\perp\cdot\nabla_\perp A\,\frac{\partial f}{\partial v_z}=0 \label{vla2}
\end{align}
Again, notice that the longitudinal coordinate $z$ plays no role in the dynamics and thus it can be integrated out to yield the equations \eqref{PCS2D-1}-\eqref{vla}.

\end{document}